\documentclass[11pt]{amsart}
\usepackage{epsfig, graphicx, color, amsmath, amsthm, amssymb}%, pst-all, pst-3dplot} %% note that pst-3dplot is not on the arxiv and needs to be included with submission
\usepackage{cite} % combines citations, e.g. [1-3] instead of [1,2,3] -- doesn't work with revtex
\input{qcircuit}

\usepackage{diagrams}
\def\ket #1{\vert #1\rangle}
\def\bra #1{\langle #1\vert}

\def\abs #1{\lvert #1\rvert}
\newcommand{\beq}{\begin{equation}}
\newcommand{\eeq}{\end{equation}}
\newcommand{\binomial}[2]{\ensuremath{\left(\begin{smallmatrix}#1 \\ #2 \end{smallmatrix}\right)}}
\newcommand\pr{{\bf {P}}}

\newcommand{\comment}[1]{\emph{\color{red}Comment:\color{black} #1}} % use for simply removing comments
\newlength{\commentslength}
\newcommand{\comments}[1]{
\hspace{-2\parindent}
\addtolength{\commentslength}{-\commentslength}
\addtolength{\commentslength}{\linewidth}
\addtolength{\commentslength}{-\parindent}
\fcolorbox{red}{white}{\smallskip\begin{minipage}[c]{\commentslength}
\emph{Comments:}\begin{itemize}#1\end{itemize}\end{minipage}}\bigskip
}
\renewcommand{\comment}[1]{}\renewcommand{\comments}[1]{}

\theoremstyle{plain} %amsart
\newtheoremstyle{note}{}{}{\slshape}{}{\bfseries}{.}{ }{}
\theoremstyle{remark} %amsart defines plain (for Theorem, Lemma, Corollary, Proposition, Conjecture, Criterion, Algorithm), definition (for Definition, Condition, Problem, Example), and remark (for Remark, Note, Notation, Claim, Summary, Acknowledgment, Case, Conclusion) theoremstyles
\newtheorem{remark}{Remark}
\theoremstyle{definition} %amsart

\DeclareMathOperator{\poly}{\operatorname{poly}}

\begin{document}

\author{Ben W. Reichardt}
\address{{EECS} Department, Computer Science Division, University of California, Berkeley, California 94720}
\thanks{Research supported in part by NSF ITR Grant CCR-0121555, and ARO Grant DAAD 19-03-1-0082.}
\email{breic@cs.berkeley.edu}

\title{Postselection threshold against biased noise}

\begin{abstract}
The highest current estimates for the amount of noise a quantum computer can tolerate are based on fault-tolerance schemes relying heavily on postselecting on no detected errors.  However, there has been no proof that these schemes give even a positive tolerable noise threshold.  A technique to prove a positive threshold, for probabilistic noise models, is presented.  The main idea is to maintain strong control over the distribution of errors in the quantum state at all times.  This distribution has correlations which conceivably could grow out of control with postselection.  But in fact, the error distribution can be written as a mixture of nearby distributions each satisfying strong independence properties, so there are no correlations for postselection to amplify.
\end{abstract}
 
\maketitle

\section{Introduction} \label{s:introduction}

The value of the quantum noise threshold, or maximum tolerable gate error rate allowing reliable quantum computation, together with the overhead required to attain it, are of considerable experimental interest.  These two parameters roughly determine how hard it is to build a useful quantum computer.  The original threshold proofs of Aharonov and Ben-Or \cite{AharonovBenOr99}, and of Kitaev \cite{Kitaev96b} showed that there exists some positive noise threshold, without giving any numerical lower bound.  Recently, though, there has been dramatic progress on two fronts of the fault-tolerance problem.  

First, there has been substantial work on estimating noise thresholds, based on simulations and heuristic analytical models, with differently-optimized fault-tolerance schemes and in different settings \cite{SvoreTerhalDiVincenzo04,Steane03,Reichardt04,SvoreCrossChuangAho}.  
In particular, Knill has recently constructed a novel fault-tolerance scheme based on very efficient distance-two codes \cite{Knill05}.  Being of distance two, his codes allow for error detection, not correction, and the scheme uses extensive postselection on no detected errors -- i.e., on detecting an error, the enclosing subroutine is restarted.  This leads to an enormous overhead at high error rates, limiting practicality.  However, Knill has estimated that the threshold for his scheme is perhaps as high as 3-6\% (independent depolarizing noise in a nonlocal gate model), a breakthrough.  (Knill also gives schemes using less postselection, and thus having more reasonable overhead but tolerating less error, too.)

Second, new, more efficient proof techniques have given explicit noise threshold lower bounds, and have shown the existence of noise thresholds for more fault-tolerance schemes and more error models \cite{AliferisGottesmanPreskill05,Reichardt05distancethree, AharonovKitaevPreskill05,AliferisTerhal05,Reichardt06Golay}.  

Despite Knill's high noise threshold estimate, though, it was not known if his scheme gave any positive threshold.  Postselection is a key factor allowing computation in the face of high error rates, but even the recent threshold proof techniques do not accommodate it, being seemingly limited to more standard threshold schemes based on error correction.  We here prove the existence of a positive constant noise threshold for a postselection-based fault-tolerance scheme.  

The intuitive problem for proving a threshold with postselection is possible negative correlations between logical errors (on the encoded state) and bit errors (away from the encoding).  (The state of the system, a distribution over pure states, can be specified by the ideal state plus a probability distribution of errors.)  For example, say in trying to prepare the $N$-bit encoded/logical state $\psi_L$, we get a logical error, $(E\psi)_L$, with some small probability.  Now postselect on no bit errors.  The good case $\psi_L$ survives with probability at least $(1-\eta)^N$ if the bit error rate is $\leq \eta$.  But if we lack any lower bounds on the bit error rate in $(E\psi)_L$, then it possible that the logical error survives with probability one, becoming exponentially more likely after renormalizing the probability distribution.\footnote{If $\psi_L$ is encoded with $k$ levels of concatenation of an $n$-bit, $t$-error-correcting code, so $N = n^k$, then the probability of $(E \psi)_L$ should be $\sim\!(c \eta)^{(t+1)^k}$ for $c$ some constant determining the threshold for improvement.  But the renormalization penalty of $\sim\!(1-\eta)^{n^k}$ overwhelms this advantage.}

If we could prove that physical errors were completely uncorrelated from logical errors -- i.e., that errors within the codespace were independent of errors going outside the codespace, so $\psi_L$ and $(E \psi)_L$ had identical bit error rates -- then the above-described problem could never occur.  Postselecting on no bit errors would improve bit reliability without affecting the distribution of logical errors.  However, this is certainly not the case.  The true error distribution has all sorts of correlations, both between different code-concatenation levels (so postselecting on no bit errors can increase the probability of logical errors, as above), and between different code blocks (so postselection in one part of the computer can harm the state of the rest of the computer).

In fact, though, the true error distribution can be written as a \emph{mixture} of error distributions which have uncorrelated errors.  By itself, that is a trivial statement, as every probability distribution can be so written -- the set of distributions is a simplex whose (deterministic) vertices have strong independence properties.  However, the mixture can be written just over nice error distributions, in which errors are not only independent, but also bounded in probability.  As we carry out the analysis, then, at every step we simply condition on a certain nice error distribution from this mixture -- it doesn't matter which one!  After implementing, say, a logical CNOT gate, the error distribution loses its independence properties, but it can again be rewritten as a mixture of distributions with bounded-probability independent errors -- this strong inductive hypothesis is restored.  

Rewriting probability distributions with small correlations as mixtures of probability distributions with bounded-probability independent events is the main technical tool of this paper.  The Mixing Lemma tells us exactly when a distribution $\pr[\cdot]$, with correlations between $n$ events, can be rewritten as a mixture of nice distributions in which those events are independent:

A point $(q_1, \ldots, q_n) \in [0,1]^n$ corresponds to a bitwise-independent distribution over $\{0,1\}^n$, in which the probability of $x$ is $\prod_{i=1}^n q_i^{x_i} (1-q_i)^{1-x_i}$.  Define the lattice ordering $y \preccurlyeq x$ for $x,y \in \{0,1\}^n$ if considered as indicators for subsets of $[n]$, $x \subseteq y$.  
\theoremstyle{theorem}
\newtheorem*{mixinglemma}{Mixing Lemma}
\begin{mixinglemma}
The convex hull, in the space of distributions over $n$-bit strings, of the $2^n$ bitwise-independent distributions $\{0,p_1\}\times\{0,p_2\}\times\cdots\times\{0,p_n\}$ is given exactly by those $\pr[\cdot]$ satisfying the inequalities, for each $x \in \{0,1\}^n$: 
\beq \label{e:mixinglemma}
\sum_{y \preccurlyeq x} (-1)^{\abs{x\oplus y}} \frac{\pr[\{z \preccurlyeq y\}]}{p(\{z \preccurlyeq y\})} \geq 0 \enspace ,
\eeq
where $p(\{z \preccurlyeq y\}) = \prod_{i=1}^n \delta_{y_i,1} p_i$, i.e., the probability of $\{z : z \preccurlyeq y\}$ in the distribution $(p_1,\ldots,p_n)$.
\end{mixinglemma}
\noindent  Note that this key lemma is completely classical, and so therefore is the essence of our argument.  The lemma's proof is deferred to Sec.~\ref{s:mixinglemmaproof}.

We illustrate the mixing technique in this extended abstract by applying it to a simple toy problem: fault-tolerance for CSS-type stabilizer operations against bit-flip errors, using the concatenated two-bit repetition code with a postselection-based scheme.  The technique generalizes further, to full universality with arbitrary Pauli errors, but most of the key insights already appear from considering just this simple example.  (Section~\ref{s:extensions} briefly describes the tricks used to extend the technique.)

We prove that there exists a constant positive threshold for this postselection-based scheme.  Computing an explicit numerical lower bound reduces to maximizing a certain function, the ratio of two quadratic polynomials, under certain bounds.  Solving for this maximum may be feasible, but is an open problem.  It is therefore unknown whether the threshold bounds derived rigorously by this method will compare favorably with Knill's high postselection threshold estimates (or even with proven threshold lower bounds for schemes without postselection), but of course that is the hope and expectation.  

\subsection*{Relation to previous work}

\begin{figure*}%figure* for wide figure
\begin{center}
\includegraphics*[bb=36.500961 37.000977 282.50681 169.00412,scale=.5]{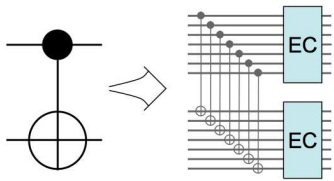}
\end{center}
\caption{To compile an ideal circuit into a fault-tolerant circuit, encode each qubit into a quantum error-correcting code and replace each ideal CNOT gate by transversal physical CNOT gates followed by error correction on each block.  Similar rules are required for the other ideal operations.  Beneath the noise threshold, some concept of ``effective reliability" of the encoded CNOT gate should improve; repeatedly applying the substitution rule gives arbitrary improvement.} \label{f:cnotsubstitution}
\end{figure*}

In 1996, Shor showed how to simulate an $N$-gate ideal quantum circuit using a physical circuit with a gate error rate of $1/\poly(\log N)$, by encoding each qubit into a $\poly(\log N)$-sized quantum error-correcting code and computing on the encoded data \cite{Shor96}.  For example, as shown in Fig.~\ref{f:cnotsubstitution}, compile each CNOT gate in the ideal circuit into transversal physical CNOT gates on the code blocks (bit 1 to 1, 2 to 2, etc.) followed by (faulty) error correction of each block to keep errors under control.  The limiting factor here is the encoding step; if we were given encoded qubits for free, with only bitwise-independent errors, then quantum fault-tolerance would be very similar to the classical fault-tolerance scheme of Von Neumann using a long repetition code $0 \mapsto 0^M$, $1 \mapsto 1^M$ \cite{VonNeumann,Knill03erasure}.  However, the quantum states for large codes are highly entangled -- for example $\ket{0}+\ket{1} \mapsto \ket{0^M} + \ket{1^M}$, a cat state -- and we can't assume that they can be prepared with bitwise-independent errors.  Aharonov and Ben-Or (AB) \cite{AharonovBenOr99} and Kitaev \cite{Kitaev96b} realized that a bootstrapping procedure based on repeatedly concatenating a constant-sized code -- and repeatedly applying the substitution rule of Fig.~\ref{f:cnotsubstitution} -- could get around this problem, and gave independent proofs of a positive \emph{constant} tolerable noise threshold.  

AB's threshold proof can be reformulated to rely on ``1-goodness."  Roughly, define a code block to be 1-good if it has at most one subblock which is not itself 1-good.  For the CNOT substitution rule of Fig.~\ref{f:cnotsubstitution}, if the two input blocks are both 1-good and at most one error occurs within the block, then the output blocks will be 1-good.  This is provided the code has distance seven or higher, for then the three total errors (one from each input block, and one during the CNOT) can be corrected in the proper direction.  Thus two errors occurring during the CNOT implementation is the bad event, so the error rate drops quadratically, giving a positive threshold.  

AB's proof can be made to work for the repeated concatenation of codes of distance-five or higher, but does not work for distance-three codes.  Reichardt \cite{Reichardt05distancethree} extends the proof to work for concatenated distance-three codes by using a stronger induction assumption.  (Aliferis, Gottesman and Preskill independently proved a threshold for concatenated distance-three codes \cite{AliferisGottesmanPreskill05}.)  In a ``1-well" block, not only is there at most one bad subblock, but also the \emph{probability} of a bad subblock is small.  The proof therefore relies on controlling the probability distribution of errors in the system as the computation progresses.  

In this paper, we similarly control the probability distribution of errors in the system, but using an even stronger induction assumption.  Whereas in Ref.~\cite{Reichardt05distancethree} it sufficed to control the errors within the ``well" blocks (and have no control over errors within bad blocks), here we need strong control over errors even within logically erroneous blocks in order to prevent postselection on no detected errors, and the subsequent renormalization, from amplifying correlations.

This paper proves a noise threshold for concatenated distance-two, error-detecting codes.  It can therefore can be seen as part of a progression in the code size for which we can prove thresholds.  The more interesting progression, though, is in the efficiency of analysis techniques.  AB and Kitaev proved the existence of some positive threshold, but proving any positive threshold at all for distance-three codes required a more efficient analysis, with less slack.  Thus the first explicit numerical lower bounds on the threshold (aside from thresholds for erasure noise \cite{KnillLaflammeMilburn00, Knill03erasure}) were derived by Aliferis, Gottesman and Preskill (AGP) and Reichardt in Refs.~\cite{AliferisGottesmanPreskill05, Reichardt05distancethree}.  We do not prove any numerical threshold lower bounds in this paper, and in any case simply trying to optimize the threshold overlooks the equally important overhead parameter.  Overhead may make schemes based on postselection, instead of error correction, uncompetitive and impractical, despite the possibly higher threshold.  However -- very speculatively -- perhaps the technique developed here, by offering even stronger control of errors in the system, will lead to more efficient analysis even of schemes which do not use postselection.

\begin{figure*}%figure* for wide figure
\begin{center}
\includegraphics*[bb=88 42 450 306,scale=.37]{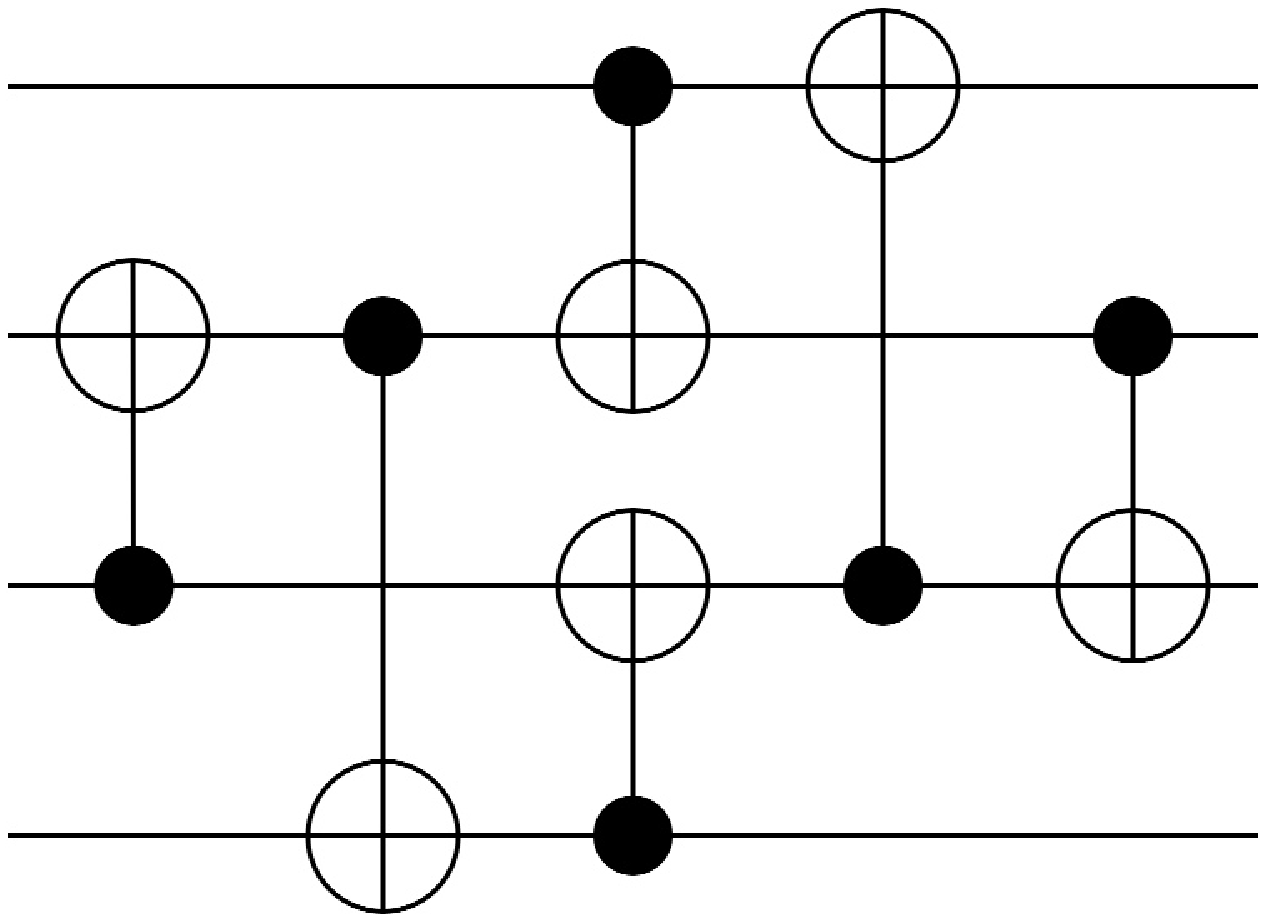} $\qquad$
\includegraphics*[bb=48 45 418 326,scale=.37]{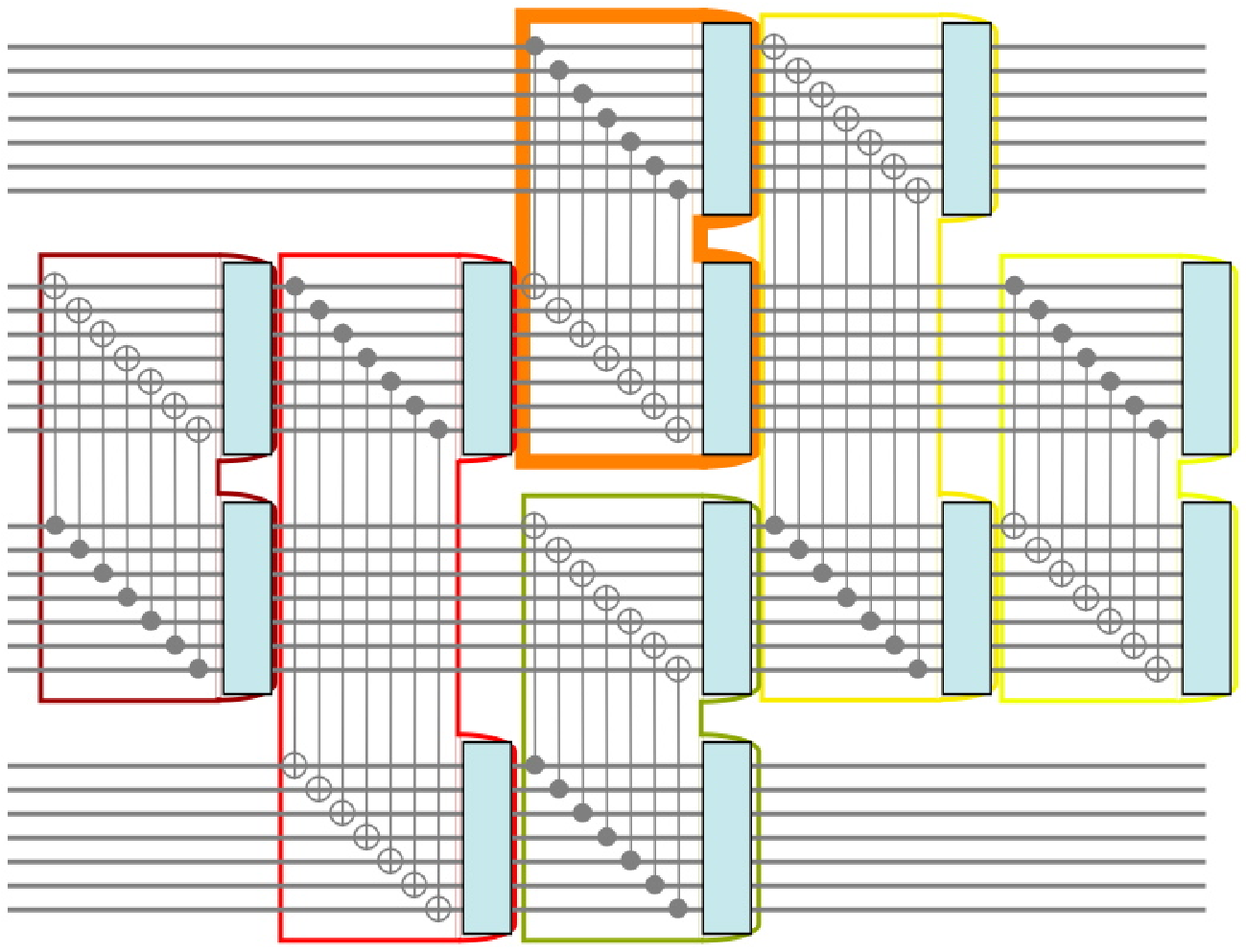} $\quad$
\includegraphics*[bb=48 45 418 326,scale=.37]{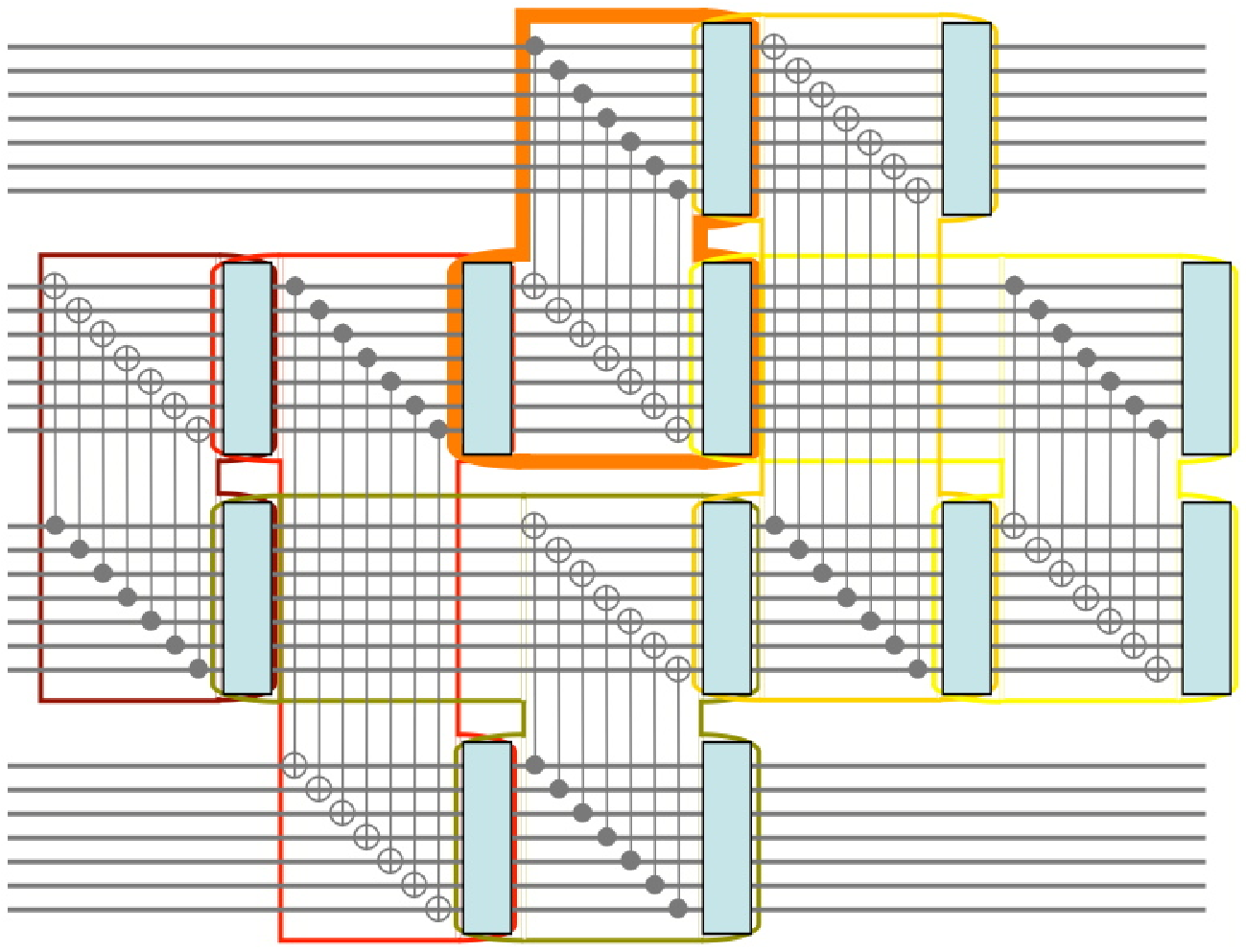}
\end{center}
\caption{Left: A certain ideal circuit using CNOT gates.  Center: The same circuit compiled with one application of the substitution rule of Fig.~\ref{f:cnotsubstitution}.  The ``rectangles," of physical gates corresponding to an ideal CNOT, are highlighted.  Right: 
Used in the proof of Aliferis, Gottesman and Preskill \cite{AliferisGottesmanPreskill05}, ``extended rectangles" -- highlighted -- overlap on the leading error correction.} \label{f:idealcircuit}
\end{figure*}

The proof technique of Ref.~\cite{Reichardt05distancethree}, like this one, is probabilistic, but analogies can be drawn between it and the distance-three code proof of AGP, which does not require a probabilistic error model.  AGP extend the basic units of AB's analysis to include also the previous error correction (Fig.~\ref{f:idealcircuit}).  They define a logical gate to have failed if two errors occur in the ``extended rectangle."  This is analogous to ``wellness" because every error in an input code block can be accounted for by some error in the previous error correction.  (The analogy is intuitive, but breaks down in the technical definitions.)  Unfortunately, it seems less likely that our new technique can be extended to coherent errors.  Writing the error distribution as a mixture of nice distributions is a classical idea which does not work for general quantum states.  In quantum mechanics language, this rewriting is equivalent to saying that the environment (in this case, the analyst!) can measure which element of the mixture the system is in -- but with coherent errors, that is simply not possible.

\section{Proof overview}

We introduce a simple independent bit-flip noise model.  We then give several lemmas each roughly saying that an encoded circuit element has the correct logical effect -- except for rare logical errors -- and outputs blocks with only weakly correlated errors (ready for the next logical gate).  Applying these lemmas at a high enough level of concatenation, logical errors will be vanishingly rare, so the encoded circuit accurately simulates the initial ideal circuit.

The key lemma required is for the encoded CNOT gate, which implemented naively would create strong bit error correlations across different blocks.  Preventing such correlations reduces to preparing an encoded Bell pair with bit errors independent across its two halves.  It is probably impossible to prepare such a state, but we can prepare a encoded Bell pair such that the error distribution can be rewritten as a mixture of nearby distributions in each of which errors are independent across the two halves.

\subsection{Error model}

Assume perfect preparation and measurement of $\ket{0/1}, \ket{\pm} = \tfrac{1}{\sqrt{2}}(\ket{0}\pm\ket{1})$ qubits, but noisy physical controlled-NOT (CNOT) gates.  Each physical CNOT gate applies an ideal CNOT gate, then fails probabilistically and independently with an error rate $\leq \eta_0$, giving bit flip (X) errors on one or both of the affected qubits.  (The ideal CNOT gate is defined by $\mathrm{CNOT}\ket{a,b} = \ket{a,a\oplus b}$, $a,b \in \{0,1\}$.)

In circuit diagram notation, we write
\begin{center}
\includegraphics[scale=.85]{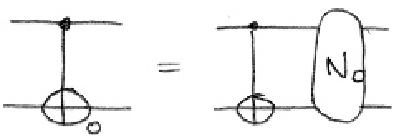}
\end{center}
Here, the left CNOT$_0$ is a physical, noisy CNOT gate, while the right CNOT is ideal.  The circled $N_0$ denotes introduction of IX, XI or XX errors with total probability at most $\eta_0$.

\subsection{Goal}

Fault tolerance is concerned with simulating ideal circuits using unreliable components.  Say we have an inputless ideal circuit $\mathcal{C}$ which merely prepares $\ket{0/1}$ and $\ket{\pm}$ qubits and applies CNOT gates to output some quantum state $\ket{\psi}$.  We construct a fault-tolerant version of $\mathcal{C}$, $\mathrm{FT}\mathcal{C}$, by computing on top of the repeatedly-concatenated two-bit repetition code.  The two-bit repetition code maps $0$ to $00$ and $1$ to $11$, and detects one bit-flip error.  Concatenated on itself $k$ times, it becomes the $2^k$-bit repetition code, mapping $b$ to $b^{2^k}$ for $b \in \{0,1\}$.

By assumption, $\ket{0}_k = \ket{0^{2^k}}$ and $\ket{1}_k = \ket{1^{2^k}}$ can be prepared perfectly.  We need to show how to prepare reliably $\ket{+}_k = \tfrac{1}{\sqrt{2}}(\ket{0}_k + \ket{1}_k) = \tfrac{1}{\sqrt{2}}(\ket{0^{2^k}}+\ket{1^{2^k}})$ (a $2^k$-bit GHZ or cat state) and how reliably to apply encoded CNOT gates.  

What does it mean to do these operations ``reliably?"  
Denote by \raisebox{-1.3ex}{\includegraphics[scale=.75]{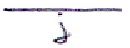}} a block of $2^j$ qubits.  Define a noisy encoding operator $\tilde{\mathcal{E}}_j$ recursively by \\
\begin{center}
\includegraphics[scale=.85]{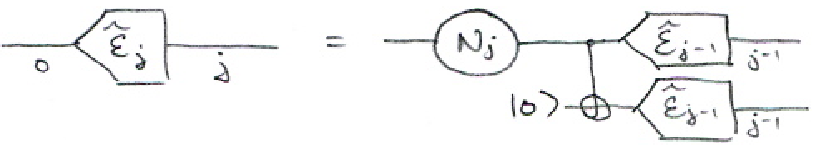} \\
\includegraphics[scale=.85]{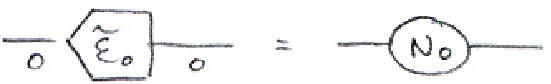} ,
\end{center}
where the circled $N_j$ means independent introduction of a bit-flip error with probability $\leq \eta_j = (c \eta_0)^{2^j}$ (some constant $c$).  $\tilde{\mathcal{E}}_j$ is not a physical operation, but is useful in our analysis.  

Reliable preparation of $\ket{+}_k$ means preparing $\tilde{\mathcal{E}}_j(\ket{+})$:
\begin{equation} \label{e:encplus}
\includegraphics[scale=.85]{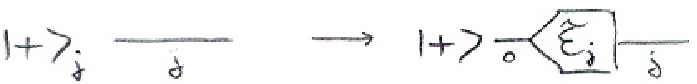}
\end{equation}
That is, noisy preparation of $\ket{+}_k$ should be the same as ideal preparation of $\ket{+}$, followed by a noisy encoding operator.  Here we write an arrow since error correlations mean we cannot enforce equality.  Our procedure for preparing $\ket{+}_k$ will produce a distribution over states which can be written as a mixture of noisy encodings of $\ket{+}$ with differing, but bounded, error parameters.  

Reliable application of a CNOT$_j$ gate means that we can commute noisy encoding operators past the encoded CNOT gate:
\begin{equation} \label{e:cnotcommute}
\includegraphics[scale=.85]{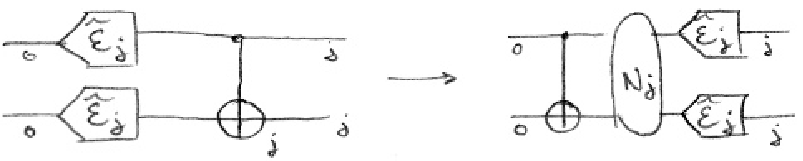}
\end{equation}
Once again, the left-hand side will be a mixture of diagrams of the type appearing on the right, all with bounded error parameters.  The right-hand CNOT gate is ideal, whereas the left-hand CNOT$_j$ indicates some implementation of an encoded gate.  The implementation will be specified below.

The simulating circuit, $\mathrm{FT}\mathcal{C}$, takes every preparation of $\ket{\phi}$ in $\mathcal{C}$ ($\phi \in \{0,1,+,-\}$) and replaces it with preparation of $\ket{\phi}_k$, and replaces every ideal CNOT in $\mathcal{C}$ with CNOT$_k$.  To analyze $\mathrm{FT}\mathcal{C}$, one repeatedly applies the above relationships to introduce noisy encoding operators $\tilde{\mathcal{E}}_k$ and then commute them past the CNOT$_k$s to the end of the circuit.  One ends up with a mixture of diagrams, each looking like the ideal $\mathcal{C}$ with noise locations $N_k$ interspersed, and noisy encoding operators applied to the output qubits.\footnote{This formalism of commuting encoding operators through the circuit, is similar to the commutative diagrams (with decoding operators) used in Refs.~\cite{Reichardt05distancethree,AliferisGottesmanPreskill05} to define logical success or failure of an encoded gate.  Here, the noisy encoding operators do not commute past perfectly, for we have to take a probabilistic mixture of diagrams on the right-hand side.}  This is our final goal; provided $k$ is large enough, so $\eta_k$ small enough (for $\eta_0 < 1/c$), it is unlikely that any of the errors $N_k$ actually occur, so we have a reliable simulation of $\mathcal{C}$.

More accurately, we want to guarantee that with high probability measurements at the end of $\mathrm{FT}\mathcal{C}$ give the same classical result as measurements at the end of $\mathcal{C}$.  It is straightforward to implement measurements; e.g., \raisebox{-1ex}{\includegraphics[scale=.75]{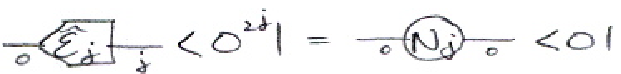}}.  The more important extensions, beyond this toy error model and to full universality, are discussed in Sec.~\ref{s:extensions}.

\subsection{Reliable preparation of $\ket{+}_j$}

The proof of Eqs.~\eqref{e:encplus},\eqref{e:cnotcommute} is by induction.  The base cases, $j=0$, are immediate, by definition of the error model.  

We implement reliable preparation of $\ket{+}_j$ as a CNOT$_{j-1}$ from $\ket{+}_{j-1}$ into $\ket{0}_{j-1}$:
\def\incl #1#2{\makebox{\raisebox{#1}{\includegraphics[scale=.8]{#2}}}}
\begin{eqnarray*}
\incl{-1.5ex}{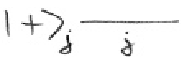} &=& \incl{-2.8ex}{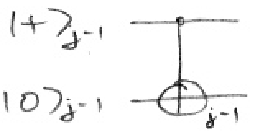} \\
&\rightarrow& \incl{-2.5ex}{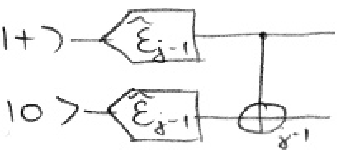} \\
&\rightarrow& \incl{-2ex}{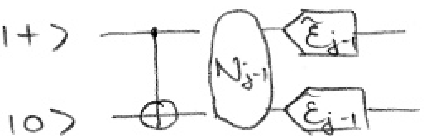} \\
&=& \makebox{\incl{-2ex}{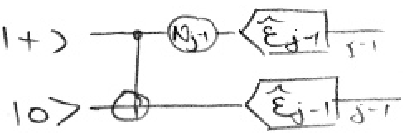}} \\
&=& \incl{-1.5ex}{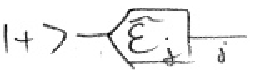}
\end{eqnarray*}
Here, the second and third lines follow from the level-$(j-1)$ versions of Eqs.~\eqref{e:encplus} and~\eqref{e:cnotcommute}, respectively.  For the fourth line: Flipping both bits has no effect on $\tfrac{1}{\sqrt{2}}(\ket{00}+\ket{11})$, so XX is equivalent to II (no error) and IX is equivalent to XI.  Thus set the probability of an error on bit two to zero, trivially independent of errors on bit one.  The last equality is by definition of a noisy encoder $\tilde{\mathcal{E}}_j$.  (This requires adjusting the constant parameters of $\tilde{\mathcal{E}}_{j-1}$.  A more careful analysis would track these parameters in order to determine the constant threshold, but to prove just the existence of a threshold, one merely has to check that the parameters stay under control.)

\subsection{CNOT gate implementation} \label{s:cnotteleport}

The fault-tolerant CNOT gate will be implemented by simultaneous teleportation and error-detection, similar to Knill's fault-tolerance scheme \cite{Knill03erasure, Knill04schemes,Knill05}.  One can verify that \\
\beq\label{e:cnotreducestobellpairs}
\makebox{\raisebox{-10ex}{\includegraphics*[bb=107.72686 213.44308 422.75421 402.2473,scale=.5]{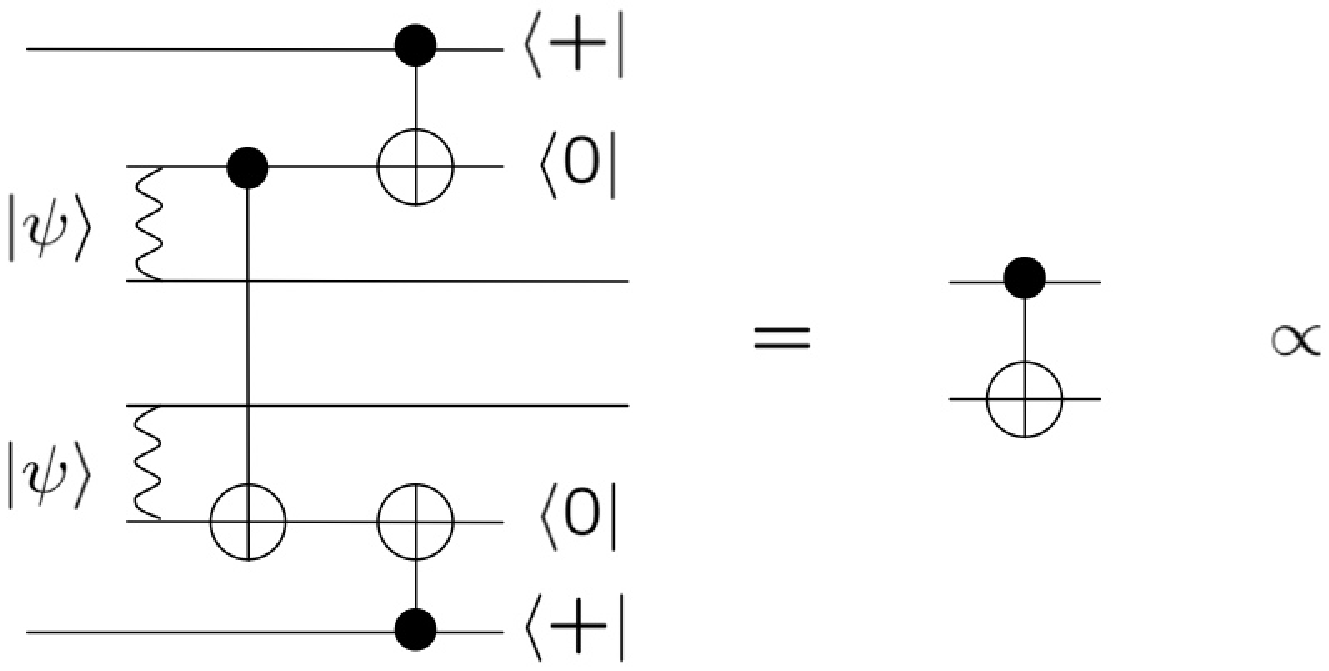}}}
\eeq
where $\ket{\psi} = \tfrac{1}{\sqrt{2}}(\ket{00}+\ket{11})$ a Bell pair, and $\bra{0}$, $\bra{+}$ denote postselected measurement of $0$ and $+$, respectively.\footnote{The success probability of this gadget is exactly $1/16$, although teleportation can be made deterministic.}

In order to implement CNOT$_j$, then, it therefore suffices to create level-$j$ encoded Bell pairs $\ket{\psi}_j$ with independent errors across the two halves (using CNOT$_{j-1}$s, $\ket{+}_{j-1}$ and $\ket{0}_{j-1}$).  For then the two CNOT$_{j-1}$s used to implement the first logical CNOT in Eq.~\eqref{e:cnotreducestobellpairs}, between the two Bell pairs, will create correlations only in blocks about to be measured anyway, not in the output blocks.  The measurement $\bra{0}$ is implemented at level $j$ by transversal measurement $\bra{0^{2^j}}$ -- i.e., postselection on no detected X errors -- while measurement $\bra{+}$ can be implemented as $\bra{+^{2^j}}$.  (We omit the details here, but this argument can be made rigorous by pushing noisy encoders through, as we did to analyze reliable preparation of $\ket{+}_j$.)  

That is, proving Eq.~\eqref{e:cnotcommute} reduces to giving a reliable preparation procedure for $\ket{\psi}_j$ satisfying:
\begin{equation} \label{e:ebellpair}
\includegraphics[scale=.95]{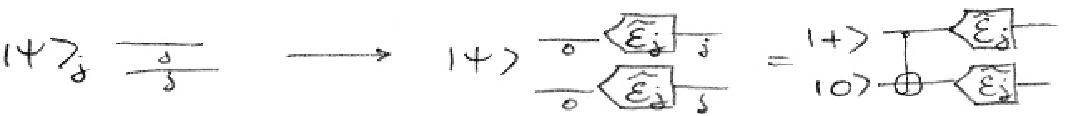}
\end{equation}

\subsection{Reliable preparation of $\ket{\psi}_j$} \label{s:reliablebellstate}

There are various ways of reliably preparing $\ket{\psi}_j$, and the choice of method has a large effect on the threshold for a particular scheme.  Here, we choose one of the simplest, shown on the left-hand side below (the boxed $\eta_{j-1}$s will be explained shortly):
\begin{eqnarray} \label{e:ebellprepproof}
\incl{-4em}{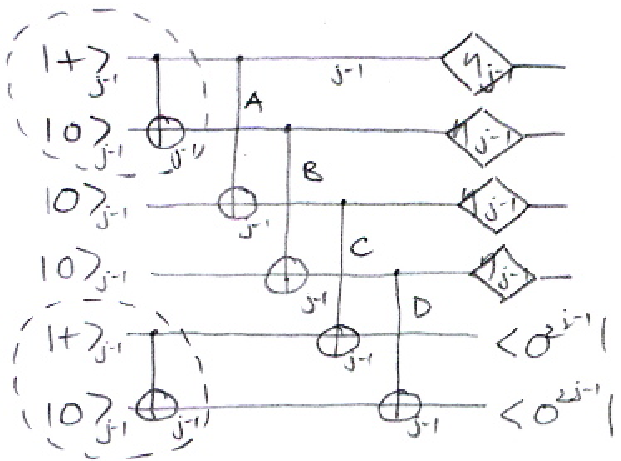} &\rightarrow& \incl{-3.7em}{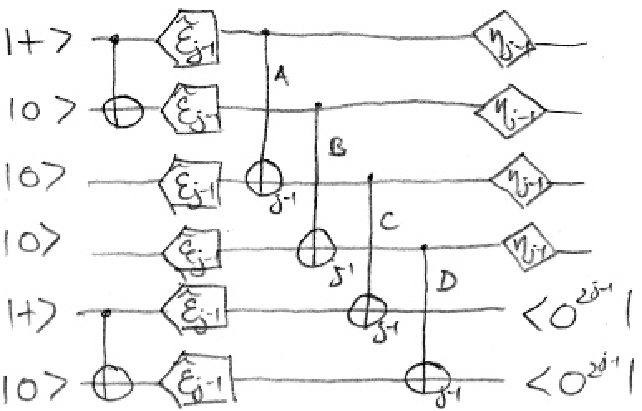} \\
&\rightarrow& \incl{-4em}{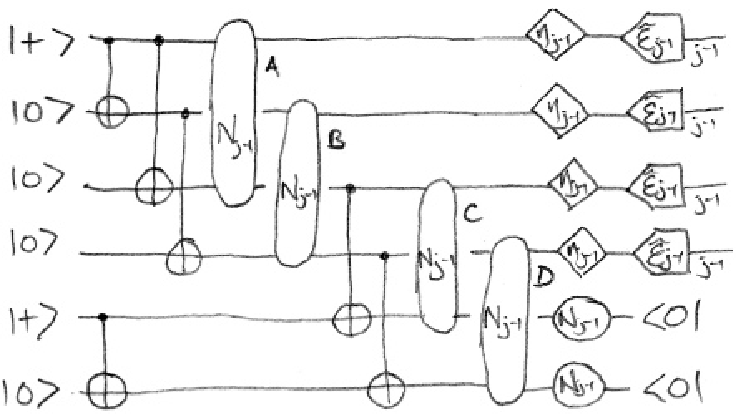} \nonumber \\ 
&\rightarrow& \incl{-2.5em}{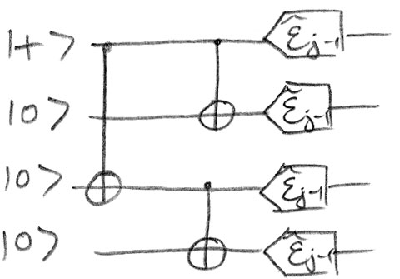} \nonumber \\
&=& \incl{-1.2em}{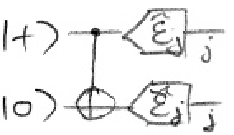} \nonumber 
\end{eqnarray}
The idea of this method is that CNOTs A and B prepare an encoded Bell pair with error correlations between its two halves.  CNOTs C and D are used to check for errors in the second half.  An error is caught if the measured block is out of the codespace, i.e., on outcomes $01$ or $10$.

In the first line above, we used Eq.~\eqref{e:encplus} twice, and in the second line used Eq.~\eqref{e:cnotcommute} at level $j-1$ as well as the measurement rule \includegraphics[scale=.75]{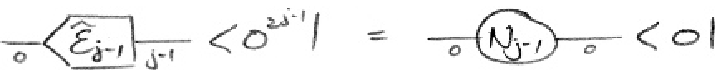}. 

\def\IIII {\mathrm{IIII}}
\def\XIII {\mathrm{XIII}}
\def\IXXX {\mathrm{IXXX}}
\def\IXII {\mathrm{IXII}}
\def\IIXI {\mathrm{IIXI}}
\def\IIIX {\mathrm{IIIX}}
\def\XXII {\mathrm{XXII}}
\def\XIXI {\mathrm{XIXI}}
\def\XIIX {\mathrm{XIIX}}
\def\IXXI {\mathrm{IXXI}}
\def\IXIX {\mathrm{IXIX}}
\def\IIXX {\mathrm{IIXX}}
\def\XXXX {\mathrm{XXXX}}
The third line, rewriting the distribution of level-$(j-1)$ errors after postselecting on acceptance as a mixture of independent error distributions, is the main step.  It can be checked directly by computing the convex hull of appropriately bounded bitwise-independent $X$ error distributions on the target encoded state: $16$ points $(0,\eta_{j-1}) \times \cdots \times (0,\eta_{j-1})$.  These points lie in only $8$ dimensions (\emph{not} $16$ dimensions labeled by $\{0,1\}^4$); since $\ket{0000}+\ket{1111}$ is unaffected by applying flipping all four bits, the different possible errors are $\IIII$ (no error), $\XIII$ (flip the first bit, equivalent to $\IXXX$), $\IXII$, $\IIXI$, $\IIIX$, $\XXII$, $\XIXI$ and $\XIIX$.\footnote{For explicit numerical calculations, linear programming software can be used to check that the convex hull of a given set of points contains a given distribution (or, all distributions satisfying certain coordinate-wise upper and lower bounds).}

The Mixing Lemma is therefore not required in this simple setting, if you are willing to get your hands dirty calculating the convex hull.  But more general error models require a larger error-detecting code and hence a larger ancilla state, and the symbolic calculation of the convex hull of a large number of points in high dimensions can be very difficult.  The Mixing Lemma gives a simple closed form for the convex hull of independent error distributions in $\{0,1\}^n$.  To illustrate its use in general, we apply it here (somewhat conservatively).

We would like to show that a distribution satisfying certain bounds lies in the convex hull of the distributions $(0,\eta_{j-1}) \times \cdots \times (0,\eta_{j-1})$, in the space $\{0,1\}^4 \mod \XXXX$.  But the Mixing Lemma only applies to points in $\{0,1\}^4$.  To apply it here, we need to \emph{linearly embed} $\{0,1\}^4/ \XXXX$ into $\{0,1\}^4$.  The simplest embedding is to evenly divide the probability mass of an error among those corresponding bit strings with minimum Hamming weight.  That is, map $\IIII$ to $0000$, $\XIII$ to $1000$, and divide the probability mass on the error $\XXII \sim \IIXX$ evenly between $1100$ and $0011$.

The Mixing Lemma gives $2^4$ inequalities to satisfy with $p_1 = p_2 = p_3 = p_4 = \eta_{j-1}$.  All except those for $x \in \{0,1\}^4$ with $\abs{x} \leq 1$ are automatic (since no probability mass has been put on strings of weight three or four).  These remaining inequalities are
%\{10000,10100,10010\}
%\{00100,10100\}
%\{00010,10010\}
\begin{gather*}
%0 & \leq & 
\begin{split}1 - \frac{1}{p_1}\pr[\{z \preccurlyeq 1000\}] - \cdots - \frac{1}{p_4}\pr[\{z \preccurlyeq 0001\}] \\ + \frac{1}{p_1 p_2}\pr[1100] + \cdots + \frac{1}{p_3 p_4} \pr[0011]  
\geq  0 \end{split}\\
\begin{split}\tfrac{1}{p_1}\pr[\{z \preccurlyeq 1000\}] - \tfrac{1}{p_1 p_2}\pr[1100] - \cdot\cdot - \tfrac{1}{p_1 p_4}\pr[1001]  \geq  0 \end{split}\\
\vdots \\
\begin{split}\tfrac{1}{p_4}\pr[\{z \preccurlyeq 0001\}] - \tfrac{1}{p_1 p_4}\pr[1001] - \cdot\cdot - \tfrac{1}{p_3 p_4}\pr[0011]  \geq  0 \end{split}
\end{gather*}
for $x = 0000, 1000, \ldots, 0001$, respectively.
It is sufficient to check instead the stronger inequalities
%\begin{subequations}
%\begin{gather}
\begin{gather*}
\frac{1}{p_1}\pr[\{\XIII,\XXII,\XIXI,\XIIX\}] + \cdots + \frac{1}{p_4}\pr[\{\IIIX,\XIIX,\IXIX,\IIXX\}] \leq 1  \nonumber %\\
\end{gather*}\begin{gather*}
\begin{split}
\frac{1}{2 p_2}\pr[\XXII] + \frac{1}{2 p_3}\pr[\XIXI] + \frac{1}{2 p_4} \pr[\XIIX] & \leq \pr[\XIII] \\
\vdots \\
\frac{1}{2 p_1}\pr[\XIIX] + \frac{1}{2 p_2}\pr[\IXIX] + \frac{1}{2 p_3} \pr[\IIXX] & \leq \pr[\IIIX]
 \enspace .
\end{split}
\end{gather*}
%\end{gather}
%\end{subequations}
The first inequality holds %for $p_1 = p_2 = p_3 = p_4 = \eta_{j-1}$ and $p_5 = \eta_j = O(\eta_{j-1}^2)$, 
because any error at all occurring, and surviving error detection, is a first-order event (in $\eta_{j-1}$).  

The other inequalities are more interesting; they roughly require that 
conditional error events be first order (but note, e.g., that $\XXII$ is the same as $\IIXX$, so the first inequality can also be written as bounding $\pr[\IIXX]$ in terms of $\pr[\XIII]$).  This is where the boxed $\eta_{j-1}$s of Eq.~\eqref{e:ebellprepproof} come in: each represents the introduction of encoded bit flip errors (application of $\mathrm{X}^{2^{j-1}}$) \emph{by the experimentalist} with probability exactly $\eta_{j-1}$.  Probabilistically introducing errors to lower-bound the right-hand side enforces these inequalities.\footnote{In probabilistically adding logical errors, one has to maintain independence with bit errors.  
As discussed in Remark~\ref{t:randomness}, %Sec.~\ref{s:extensions}, 
this is difficult -- unless physical NOT gates are perfect.  However, it can be done by only introducing the errors in your head, and tracking them with a classical computer.}

The Mixing Lemma now tells us that the embedded image of our error distribution can be rewritten as a mixture of bounded, bitwise-independent distributions, in $\{0,1\}^4$.  But the embedding is linear, and also satisfies:
\begin{enumerate}
\item An error distribution is uniquely determined by its image, i.e., each $x \in \{0,1\}^4$ corresponds to a unique error equivalence class.
\item With the same reverse map as in (1), every bitwise-independent distribution is the image of a bitwise-independent error distribution (with a possibly different division of probability mass).
\end{enumerate}
Under these conditions, the image of an error distribution lying in the convex hull of bitwise-independent distributions in $\{0,1\}^4$, will imply that the error distribution is the same convex combination of corresponding bitwise-independent error distributions in $\{0,1\}^4 / \XXXX$.  (Since this correspondence is quite clear in our simple case, we avoid elaborating the notation.  But more interesting embeddings -- e.g., with careful division of probability mass, or into longer bit strings -- can sometimes be useful.)

Deliberately introducing errors into the computation is counterintuitive, but is necessary for applying the Mixing Lemma.  For example, say that only failure locations A and C in Eq.~\eqref{e:ebellprepproof} are faulty, and the other locations are perfect; and moreover that A only fails as XX and C only fails as IX.  Then $\pr[\XIXI] > 0$ but $\pr[\XIII] = \cdots = \pr[\IIIX] = 0$ -- for acceptance, neither or both of A and C must fail.  This distribution cannot be written as a mixture of distributions with bitwise independent X errors.  

\begin{remark} \label{t:thresholdpenalty}
Introducing errors might well hurt the threshold (presuming a postselection threshold even exists without introducing errors), but probably by no more than a small constant factor.  If we are given lower bounds on CNOT gate failure rates, then it may not be necessary to deliberately introduce errors.  For example, if we are guaranteed that all physical CNOT$_0$ gates fail with the \emph{same} probability, $\leq \eta_0$, and on failure IX, XI and XX errors are equally likely, then it is not necessary to introduce any errors in constructing $\ket{\psi}_1$.\footnote{Also, this scheme can be modified so it isn't ever necessary to deliberately introduce errors, but only if one uses top-level error correction when verifying encoded Bell pairs, and error detection elsewhere.}
\end{remark}

\begin{remark}[Randomness] \label{t:randomness}
It is often useful to apply a gate (e.g., a swap gate or a Hadamard) with say probability $1/2$, in order to symmetrize a state.  This is more difficult in our encoded setting 
because noise will not be independent of whether or not the logical operation was applied.  For example, if one randomly swaps two blocks having independent errors, the output state will usually have correlations between the blocks.  This is why we did not assume a symmetrical error model, $\pr[\XIII] = \cdots = \pr[\IIIX]$.  
\end{remark}

\begin{remark}
The proof's inductive structure is:
$$
\begin{diagram}[balance,width=3em,height=2em,tight]
\ket{+}_0 & \rTo & \ket{\psi}_0 & \rTo & \ket{\psi}_1 & \longrightarrow                    & \cdots & 
\longrightarrow                   & \ket{\psi}_{j-1} & \rTo  & \ket{\psi}_j & \longrightarrow & \cdots \\
          &      & \uTo         & \ruTo & \dTo & \rotatebox{30}{$\longrightarrow$} &        & 
\rotatebox{30}{$\longrightarrow$} & \dTo         & \ruTo      & \dTo   &   \rotatebox{30}{$\longrightarrow$} & \\
          &      & CNOT_0       & & CNOT_1 &                                   & \cdots & 
                                  & CNOT_{j-1}   &       & CNOT_j   & & \cdots 
\end{diagram}
$$
(Reliable preparation of $\ket{+}_j$ is equal to reliable preparation of $\ket{\psi}_{j-1}$ for this code, so was not actually needed.)
\end{remark}

\section{Extensions and open problems} \label{s:extensions}

We briefly sketch some concerns with this model, extensions and open problems.

\subsection{Universality} \label{s:universality}

We have not shown here how to implement reliably a universal set of quantum gates; the CNOT gate, together with preparation and measurement of $\ket{0/1}$, $\ket{\pm}$, is a subset of the set of ``stabilizer operations" which are efficiently simulatable classically \cite{AaronsonGottesman04}.  The extension to universality is via the technique of magic states distillation \cite{BravyiKitaev04, Reichardt04magic, Reichardt05distancethree, Reichardt06magic} (although one needs to be careful about randomization -- see Remark~\ref{t:randomness}).  Magic states distillation lets us obtain universality at level $k$ using only level-$k$ stabilizer operations and certain unencoded noisy ancilla preparations.  

\subsection{Biased X noise model} 

This analysis can be extended to more interesting noise models, including both bit flip (X)  and phase flip (Z) errors.  The smallest quantum error-detecting code detecting both kinds of errors uses four qubits -- for example, concatenate the repetition code $b \mapsto bb$, $b \in \{0,1\}$ onto the dual repetition code $\ket{\pm} \mapsto \ket{\pm\pm}$.  The encoded Bell pair has eight qubits, and the Mixing Lemma can be applied with $n=24$ (three error possibilities, X, Z and Y$=i$XZ, for each bit) and nontrivial inequalities only for $x$ with $\abs{x} \leq 2$.  (The Mixing Lemma also generalizes to the natural lattice on $\{I, X, Y, Z\}^8$.)

By itself, the bit-flip noise model presents interesting challenges.  For example, there are likely better ways of preparing large cat states -- e.g., adding a single qubit at a time instead of doubling -- but these can be difficult to analyze.  Actually, even reliably implementing all the stabilizer operations requires tricks in the biased noise case; because the Hadamard gate sends bit flip to phase flip errors, which the repetition code does not protect against.

\subsection{Asymptotic efficiency and composition with other fault-tolerance schemes}

The error rate with this scheme drops doubly-exponentially fast in $k$ the number of levels of code concatenation, meaning $k$ must be $\poly(\log \log N)$ to reliably simulate an $N$-gate circuit $\mathcal{C}$.  The overhead is growing as $\exp(N \exp(k))$.  The overhead can be made polylogarithmic by teleporting into the first of two levels of large random codes \cite{CalderbankShor96}, following Knill in Ref.~\cite{Knill03erasure}.

Can a postselection-based fault-tolerance scheme be concatenated on top of, or below, other fault-tolerance schemes?  This changes the base error model.

\subsection{Numerics}

We have proved the existence of a constant noise threshold, but no explicit threshold lower bound.  In his simulations, Knill found that the error distribution was quite close to having independent errors \cite{Knill04analysis}.  Therefore, writing the true error distribution as a mixture of nearby distributions with independent errors has the potential 
to give good threshold lower bounds.  (See Remark~\ref{t:thresholdpenalty}.)  Calculations, with more careful tracking of parameters, are in progress.

There are many ways of optimizing the presented fault-tolerance scheme.  For example, it is probably better to verify against errors the full ancillary state used in implementing encoded CNOT gates (Sec.~\ref{s:cnotteleport}).  But rather than rediscover optimizations, it makes sense to analyze Knill's scheme, which has been optimized already using simulations.

Related questions: Can this proof method be applied to give reasonably high threshold lower bounds for fault-tolerance schemes which do not use postselection?\comment{For example, Knill's other scheme in his Nature paper.}  Might the Mixing Lemma be useful even for obtaining nonrigorous threshold estimates?  

\subsection{Local gates}

Physical constraints typically dictate that only neighboring or nearby qubits can interact with each other \cite{Gottesman00local,SvoreTerhalDiVincenzo04,MetodievCrossThakerBrownCopseyChongChuang04}.  We have however assumed that CNOT gates can be applied between arbitrary qubits.  Locality is not a particular problem for our proof technique, but may hamper postselection-based fault-tolerance schemes.

\subsection{Probabilistic noise}

We have assumed that the noise is probabilistic, with each gate failing with a Pauli error independently of the others.  Threshold results, for fault-tolerance schemes not based on postselection, exist for more general and more physically-realistic error models \cite{KnillLaflammeZurekScience98,AliferisGottesmanPreskill05,AharonovKitaevPreskill05}.  Some noise correlations can be dealt with by applying the Mixing Lemma to the physical noise itself.  However, this proof technique may be constrained to probabilistic Pauli noise models.

\section{Proof of the Mixing Lemma} \label{s:mixinglemmaproof}

Recall the lattice ordering; e.g., $110,101,111 \preccurlyeq 100$.  
For $w \in \{0,1\}^n$, let $w \cdot p$ denote the distribution $(w_1 p_1, \ldots, w_n p_n)$.  I.e., if $Z$ is drawn from $w \cdot p$, then
$
(w \cdot p)(z) \equiv \pr[Z = z] 
= \prod_{i=1}^n (w_i p_i)^{z_i} (1- w_i p_i)^{1-z_i} 
$.  
In particular, 
$$
(w \cdot p)(\{z : z \preccurlyeq y\}) = \left\{\begin{array}{ll}\Pi_{i \in y} p_i = p(\{z \preccurlyeq y\}) & \textrm{if $w \preccurlyeq y$} \\  0 & \textrm{o.w.}\end{array}\right.
$$

The convex hull of the distributions $\{w \cdot p : w \in \{0,1\}^n\}$ is contained in the set specified by the simultaneous inequalities~\eqref{e:mixinglemma}.  Indeed, $w \cdot p$ satisfies all the inequalities with equality, except that for $x = w$ for which it gives $1$:
%\begin{eqnarray*}
\begin{equation*}
\sum_{y \preccurlyeq x} (-1)^{\abs{x \oplus y}} \frac{(w \cdot p)(\{z \preccurlyeq y\})}{p(\{z \preccurlyeq y\})} 
= \sum_{y : w \preccurlyeq y \preccurlyeq x} (-1)^{\abs{x \oplus y}} 
= \delta_{x,y} %\enspace ,
\end{equation*}
%\end{eqnarray*}
since necessarily $x \succcurlyeq w$ for the sum over $y$ to be nonzero, and then $\sum_{k=0}^{\abs{w}-\abs{x}} \binomial{\abs{w}-\abs{x}}{k} (-1)^k = 0$ unless $\abs{x} = \abs{w}$.

Conversely, if a distribution $\pr[\cdot]$ satisfies inequalities~\eqref{e:mixinglemma}, then it lies in the convex hull of the distributions $\{w \cdot p : w \in \{0,1\}^n\}$.  Indeed, the $w \cdot p$ coordinate of $\pr[\cdot]$ is given by the value of the left-hand side of Eq.~\eqref{e:mixinglemma} for $x=w$.  These coordinates are nonnegative, and using these coordinates recovers $\pr[\cdot]$; for all $v \in \{0,1\}^n$,
\begin{eqnarray*}
\sum_{\substack{x,y\\ y \preccurlyeq x}} (-1)^{\abs{x\oplus y}} \frac{\pr[\{z \preccurlyeq y\}]}{p(\{z \preccurlyeq y\})} (x \cdot p)(\{z \preccurlyeq v\}) 
&=& \sum_{\substack{x,y\\ y \preccurlyeq x \preccurlyeq v}} \pr[\{z \preccurlyeq y\}] (-1)^{\abs{x\oplus y}} \frac{\prod_{i \in v} p_i}{\prod_{i \in y} p_i} 
\\ &=& \pr[\{z \preccurlyeq v\}] \enspace ,
\end{eqnarray*}
since again the sum over $x$ is zero unless $y=v$.
(The values $\pr[\{z \preccurlyeq v\}]$ for different $v$ characterize $\pr[\cdot]$.)
\qed

%\bibliographystyle{bibtex/hunsrt}
%\bibliography{tun}

\end{document}